\documentclass[12pt]{article}
\usepackage{epsfig}
\usepackage{array}
\makeatletter
\def\alt{\mathrel{\mathpalette\gl@align<}}
\def\agt{\mathrel{\mathpalette\gl@align>}}
\def\gl@align#1#2{\lower.6ex\vbox{\baselineskip\z@skip\lineskip\z@
\ialign{$\m@th#1\hfil##\hfil$\crcr#2\crcr\sim\crcr}}}
\makeatother

\begin{document}
\begin{flushright}
{\tt hep-ph/0512171}\\
MIFP-05-36 \\
December, 2005 \\
\end{flushright}
\vspace*{2cm}
\begin{center}
{\baselineskip 25pt \large{\bf
Properties of Fermion Mixings in Intersecting D-brane Models
} \\

}

\vspace{1cm}

{\large
Bhaskar Dutta and
Yukihiro Mimura
} \vspace{.5cm}

{
\it Department of Physics, Texas A\&M University,
College Station, TX 77843-4242, USA
}
\vspace{.5cm}

\vspace{1.5cm}
{\bf Abstract}
\end{center}

We consider the Yukawa couplings for quarks and leptons in the context of
Pati-Salam model using intersecting D-brane
models where the Yukawa coupling matrices are rank one in a simple choice of
family replication. The
 CKM mixings can be
explained by perturbing the rank 1 matrix using higher order terms
involving new Higgs fields available in the model.  We show that the
near bi-large neutrino mixing angles can be naturally explained,
choosing  the light neutrino mass matrix to be type II seesaw
dominant. The predicted value of $U_{e3}$ is  in the range $\simeq
0.05-0.15$.
In the quark sector, $V_{cb}$ is naturally close to the strange/bottom quark
mass ratio
and we obtain an approximate relation $V_{ub} V_{cb} \simeq ({m_s}/{m_b})^2 V_{us}\,$.
The geometrical interpretations of the neutrino mixings are also
discussed.

\thispagestyle{empty}

\bigskip
\newpage

\addtocounter{page}{-1}

\section{Introduction}
\baselineskip 20pt

Understanding the masses and the mixings of quarks and leptons is
one of the most important issues in particle physics.
In the quark sector, there are 6 quark masses, $(m_u,m_c,m_t)$
for up-type quarks and $(m_d,m_s,m_b)$ for down-type quarks.
For the quark mixing, there are 3 mixing angles in the CKM
(Cabibbo-Kobayashi-Maskawa) matrix, and 1 phase.
The masses are hierarchical and none of the mixing angles is large.
Although those 10 parameters are completely free in the standard model,
there may exist certain relations among the mass ratios and the mixing
angles~\cite{Fritzsch:1977za}.

 On the other hand, in the leptonic sector, there are 3 charged-lepton masses
$(m_e,m_\mu,m_\tau)$, whose hierarchical pattern is similar to
the down-type quark ones, though it is not completely same \cite{Georgi:1979df}.
Recent neutrino experiments show that neutrinos also have masses and it was
revealed~\cite{Fukuda:1998tw,Ahmad:2001an} that the two neutrino mixing angles
to explain the atmospheric and solar neutrino data are large
(especially, the best fit for atmospheric mixing is maximal),
while another mixing angle $\theta_{13}$ is small as required to satisfy the
long baseline neutrino data \cite{Apollonio:1997xe}.
In fact, these hierarchical mass patterns and this combination of small
and large mixing angles may be a key issue to select models beyond the standard
model
and to explain the origin of flavors.
In the framework of the standard model, there is no relation between the quark
sector
and the leptonic sector.
It is discussed whether the quark and lepton masses and mixings can be related
in an unification pictures \cite{Fukuyama:2002ch}.

Even if the unified gauge models are considered,
the Yukawa couplings are fundamental parameters in four-dimensional field theory.
In that case, the existence of more fundamental theories are expected to describe
 the variety of quark and lepton masses and mixings.
String theory is a most promising candidate to describe particle field theories
as an effective theory, as well as quantum gravity.
String theory is attractive because all the parameters can be calculated
from a few fundamental parameters.
But there has been no clear answer on how to derive the standard model
in string theory since the selection of vacua may be a non-perturbative phenomena.
However, the non-perturbative aspects of  string theories can be discussed, once the
D-branes were formulated 
\cite{Polchinski:1995mt}.
Indeed, the intersecting D-branes \cite{Berkooz:1996km,Blumenhagen:2000wh}
are interesting approaches to construct the standard model.
The $N$ stack of D-branes can form $U(N)$ gauge fields,
and at the intersection between the $N$ stack and $M$ stack of D-branes,
a massless chiral fermion belonging to $(N,\bar M)$ bi-fundamental
representation can appear.
Such a situation is very attractive to obtain quark and lepton fields
not only in the standard model but also in the models where the  gauge
group is given as direct group such as Pati-Salam model~\cite{Pati:1974yy}.
In addition to the realization of the particle representation
and the gauge groups of the standard-like models,
the Yukawa couplings are calculable in the intersecting D-brane
models~\cite{Cremades:2003qj,Cvetic:2003ch,Higaki:2005ie}.
The couplings are described as $e^{-k A}$ naively by the triangle area $A$
formed by the three intersecting points. The presence of the exponential factors
can be utilized to achieve
the hierarchical pattern of fermion masses.

Since the intersecting D-brane models have potentials to explain
the pattern of fermion masses and mixings, many people
have constructed various intersecting D-brane models
\cite{Cvetic:2001tj,Kokorelis:2002ip,Cvetic:2004ui,Marchesano:2004yq}.
One interesting issue is that in the simple models,
Yukawa matrices are written as factorized form $y_{ij} = x^L_i x^R_j$
\cite{Cremades:2003qj,Chamoun:2003pf}.
This originates from a geometrical reason that
the left- and right-handed fermions are replicated at the intersecting
points on the different tori, and the Yukawa couplings are given
as an exponential form of sum of the triangle areas.
As a result of the factorized form of Yukawa coupling,
the Yukawa matrices are rank 1, and thus only the 3rd generation fermions are
massive.
In order to construct a realistic model, this issue for
Yukawa matrices needs to be resolved and
several possibilities have been considered in the
literature~\cite{Chamoun:2003pf,Abel:2003yh,Kitazawa:2004ed}.

In this paper, we emphasize the possibility that the rank 1 Yukawa matrices
are crucial to understand the properties of fermion mixings.
The Pati-Salam model can be constructed using intersecting D-branes
with several attractive features \cite{Cvetic:2004ui}.
This model has left-right gauge symmetry $SU(2)_L \times SU(2)_R$ and
an up-down symmetry is exhibited if there is only one Higgs bidoublets.
This up-down symmetry must be broken since the up- and the down-type
quark masses have different hierarchical pattern and the CKM matrix
is not an identity matrix.
Consequently, new Higgs fields must be introduced to break up-down symmetry.
The extra Higgs fields are also needed to raise the rank of the Yukawa matrices.
In fact, there are extra Higgs fields at the intersection
between visible branes and hidden branes,
which is needed to satisfy the RR tadpole cancellation condition.
Such extra Higgs fields can contribute to produce the
Yukawa matrices through  higher order terms,
and hierarchies of the fermion masses can be realized.
We also study the consequences that the Yukawa couplings are given
as rank 1 matrices plus small corrections.
We will show that the observed small mixings in quark sector
can be easily realized, and in the lepton sector,
the  solar and the atmospheric mixings for neutrino oscillation,
are generically large while one other mixing is small.
%
We will also study the geometrical interpretation of the neutrino mixings.

This paper is organized as follows:
In section 2, we construct intersecting D-brane models
with the Pati-Salam gauge symmetry.
In section 3, the Yukawa matrices in the intersecting
D-brane models are studied and we discuss how the
almost rank 1 Yukawa matrices are realized.
In section 4, we show the consequences of the fact that
Yukawa matrices are almost rank 1 matrix.
In section 5, we will see that the observed properties
of neutrino mixings can be interpreted in geometrical way.
The section 6 is devoted to  conclusions and discussions.

\section{Pati-Salam like model from intersecting D-branes}

In this section, we will briefly discuss the construction of a model,
in the type IIA orientifolds on $T^6/Z_2\times
Z_2$ with intersecting D6-branes \cite{Blumenhagen:2005mu}.
The supersymmetric Pati-Salam models with gauge symmetry
$U(4)_c \times U(2)_L \times U(2)_R \times G_{h}$ are constructed in
the Ref.~\cite{Cvetic:2004ui}.
There are D6$_{a}$-brane for $U(4)_c$, D6$_b$-brane for $U(2)_L$,
D6$_c$-brane for $U(2)_R$.
Extra branes are needed to cancel RR tadpole.
We call such extra branes D6$_{1,2}$-branes which provide
hidden $USp$ gauge groups.

At the intersection between the D6$_a$-brane and the D6$_b$-brane,
for example, open strings can stretch and chiral
fermions belonging to bi-fundamental representation
$(\bf{4,{\bar 2},1})$ can appear as a zero mode which corresponds to
the left-handed matter fields.
The right-handed matter fields $(\bf{{\bar 4},1,2})$
can be located at the intersection between the D6$_a$-brane and the D6$_c$-brane.
The Higgs bidoublet $({\bf 1,2,{\bar 2}})$ is at the intersection
between the D6$_b$-brane and the D6$_c$-brane.
The family is replicated when the six dimensions are compactified
to the torus $T^6 = T^2 \times T^2 \times T^2$.
The family number is given by the intersecting number
\begin{equation}
I_{\alpha\beta} = \prod_{i=1}^3 (n_\alpha^i m_\beta^i - m_\alpha^i n_\beta^i),
\end{equation}
using wrapping numbers $(n_\alpha^i,m_\alpha^i)$ for each torus $(i=1,2,3)$,
which specifies that the D6$_\alpha$-branes are
stretching over our three-dimensional space.
An orientifold, which is needed to have negative contribution
to the vacuum energy, can be constructed by discrete
transformation with world sheet parity.
The orientifold image of $\alpha$ brane is denoted as $\alpha^\prime$.
The wrapping numbers of the $\alpha^\prime$ are given as $(n_\alpha^i,-m_\alpha^i)$.
%
%
In order to obtain odd numbers of chiral families in this model, we need
 one tilted torus, and for the tilted torus we have
 $\tilde m_\alpha^i = m_\alpha^i + \frac12 n_\alpha^i$
and the wrapping number of the orientifold image is $(n_\alpha^i,-\tilde m_\alpha^i)$.
The wrapping numbers are
constrained by the RR tadpole cancellation and the supersymmetry preserving
conditions. The wrapping numbers for the
supersymmetric Pati-Salam models with three chiral families are systematically
searched in  Ref.\cite{Cvetic:2004ui}. An
example of wrapping numbers from one of the models in  Ref.\cite{Cvetic:2004ui}
is given in  Table 1.
%
%
\begin{table}[tbp]
\center
\begin{tabular}{|c||c||c|c|c|}\hline
  & $N_\alpha$ & $(n^1_\alpha,m^1_\alpha)$ &$(n^2_\alpha,m^2_\alpha)$ &
  $(n^3_\alpha, \tilde m^3_\alpha)$ \\ \hline \hline
$a$ & $8$ & $(0,-1)$ & $(1,1)$ & $(1,1/2)$ \\ \hline
$b$ & $4$ & $(3,1)$  & $(1,0)$ & $(1,-1/2)$ \\ \hline
$c$ & $4$ & $(1,0)$  & $(1,4)$ & $(1,-1/2)$ \\ \hline \hline
$1$ & $4$ & $(0,-1)$ & $(0,1)$ & $(2,0)$ \\ \hline
$2$ & $2$ & $(0,-1)$ & $(1,0)$ & $(0,1)$ \\ \hline
\end{tabular}
\caption{An example of wrapping numbers to obtain supersymmetric
Pati-Salam model with  three chiral families. }
\end{table}
The number $N_\alpha$ in  Table 1 denotes the stack number of the $\alpha$ brane.
In the $T^6/Z_2\times Z_2$ orbifold
model, $N_\alpha$ stack of branes generates gauge symmetry $U(N_\alpha/2)$.
For the branes which parallel to the
orientifolds,
$USp(N_\alpha)$ gauge symmetry is generated. 
So the gauge symmetry for the wrapping numbers shown in Table 1 is
$U(4)_c \times U(2)_L \times U(2)_R \times USp(4)_1 \times USp(2)_2$.



\begin{table}[tbp]
\center
\begin{tabular}{|c||c||c|c|c|c||c|}\hline
sector & rep. & $Q_4$ & $Q_{2L}$ & $Q_{2R}$ & $Q^\prime$& field \\ \hline\hline
$ab$ & $(\bf{4,\bar 2,1})$ & 1 & $-1$ & 0 & 0 & $\psi$  \\ \hline
$ac$ & $({\bar{\bf{4}},\bf 1,2})$ & $-1$ & 0 & 1 & 0 & $\psi^c$  \\ \hline
$bc$ & $(\bf{1,2,{\bar 2}})$ & 0 & 1 & $-1$ & 0 & $H$  \\ \hline
$a1$ & $(\bf{4,1,1})$ & 1 & 0 & $0$ & $\pm1$ & $X$  \\ \hline
$b1$ & $(\bf{1,{2},1})$ & 0 & $1$ & $0$ & $\pm 1$ & $h$  \\ \hline
$c1$ & $(\bf{1,1,{\bar 2}})$ & 0 & 0 & $-1$ & $\pm 1$ & $h^c$  \\ \hline
\end{tabular}
\caption{ The relevant fields to construct a semi-realistic model.
The conjugate representations 
are denoted as $\bar X$, for example.}
\end{table}

We list relevant fields to construct our model in Table 2.
The $Q_4$, $Q_{2L}$, and $Q_{2R}$ are the charges for $U(1)$'s which
are subgroups of $U(4)_c$, $U(2)_{2L}$, $U(2)_{2R}$.
Those three $U(1)$ symmetries are anomalous and their gauge bosons become
massive by generalized Green-Schwarz mechanism~\cite{Aldazabal:2000dg}.
 The $Q^\prime$ denotes the $U(1)$ charge embedded in $USp(4)$.
The $USp(4)$ symmetry is broken to $U(1)$ by splitting away
the branes from the orientifolds on all three tori, which is
equivalent to Higgsing of $USp(4)$ by three antisymmetric
chiral multiplets which are massless modes \cite{Cvetic:2004nk}.
The fundamental representation of $USp(4)$ group has $\pm 1$ charges
under the $U(1)$ subgroup.
The hypercharge of the fields can be defined as
\begin{equation}
Y = T_{2R}^3 + \frac{B-L}2 + 
\frac{Q^\prime}2 \,.
\label{hypercharge}
\end{equation}
The $B-L$ is a $U(1)$  generator, diag.($1/3,1/3,1/3,-1$),
in the $SU(4)_c \rightarrow SU(3)_c \times U(1)_{B-L}$.
The $SU(4)_c \times SU(2)_R$ symmetry can be broken to
$SU(3)_c \times U(1)_{B-L} \times U(1)_R$ by the brane splitting.
The vacuum expectation value of $h^c$ 
breaks $U(1)_{R} \times U(1)_{USp}$ symmetry to $U(1)_R^\prime$. %
%
The brane splitting is equivalent to the Higgsing by adjoint fields
for $SU(4)_c$ and $SU(2)_R$ in the effective field theory,
thus Yukawa couplings for top, bottom and tau may be
almost unified since the breaking terms are higher order.


We note that the model is consistent when we consider the
 confining phase of $USp(4)$ where $h h^c$ is confined to be
the bidoublet Higgs fields.
Also, $X h$ (and $X \bar h$) can be confined to form the matter representation,
and hence, $X$ should be vector-like to maintain chiral three generations.
When the confining field $\bar X h^c$ acquires vacuum expectation values,
$SU(4)_{c} \times SU(2)_R$ is broken down to $SU(3)_c \times U(1)_Y$.
The beta functions for the $USp$ groups are negative for the model
shown in the Table 2.
The confining $USp$ gauge groups may be interesting
since it may break the supersymmetry
by gaugino condensation mechanism \cite{Cvetic:2003yd}.


Furthermore, both left- and right-handed neutrino Majorana mass terms
%
%
can be generated by the non-renormalizable interaction in the superpotential,
\begin{equation}
\frac1{M_s^3} f_{ij} \psi_i \psi_j \,[\bar X h \bar X h] +
 \frac1{M_s^3} f_{ij}^c \psi^c_i \psi^c_j \,[X h^c X h^c] \,.
\end{equation}
The $USp$ confinement can produce $SU(2)_{L,R}$ triplets with proper hypercharges,
$[\bar X h \bar X h] \sim \Lambda^3 \bar \Delta_L$ and
$[X h^c X h^c] \sim \Lambda^3 \Delta_R$,
where $\Lambda$ is a confining scale of $USp$ group.
Once $SU(2)_L$ triplet acquires a
small vacuum expectation value around sub-eV range and $SU(2)_R$ triplet
acquires a vacuum expectation value around
 the confining scale, the seesaw neutrino masses \cite{Minkowski:1977sc},
\begin{equation}
m_\nu^{\rm light} = M_L - M_\nu^D M_R^{-1} (M_\nu^D)^T,
\label{seesaw}
\end{equation}
are generated,
where $M_\nu^D$ is a neutrino Dirac mass matrix and
$M_L$ and $M_R$ are left- and right-handed neutrino Majorana mass matrices
which are proportional to $f$ and $f^c$.


\section{Yukawa couplings}

The Yukawa couplings, 
\begin{equation}
W = y_{ij} \psi_i \psi^c_j H,
\end{equation}
%
are generated from world sheet instanton corrections
and is written by using Jacobi theta function in the form \cite{Cremades:2003qj},
\begin{equation}
y_{ij} \propto \prod_{r=1}^3
\vartheta \left[
\begin{array}{c}
\delta^{(r)} \\ \phi^{(r)}
\end{array} \right](\kappa^{(r)})\,,
\quad
\vartheta \left[
\begin{array}{c}
\delta \\ \phi
\end{array} \right](\kappa) =
\sum_{\ell \in {\bf Z}} e^{-\pi \kappa (\delta +\ell)^2}
e^{2\pi i(\delta+\ell)\phi}, \label{theta-function}
\end{equation}
where $\delta^{(r)}$ characterizes intersections and shifts of the D-branes
in each torus, $\kappa^{(r)}$ represents a K\"ahler structure,
and $\phi$ is for a possible Wilson-line phase.

It is important that the Yukawa couplings are given in a factorized form.
As a result, when the family of left- and right-handed matters are
replicated in different tori as in the example given in Table 1,
the Yukawa matrices of quarks
and leptons are given as $y_{ij} \propto x_i^L x_j^R$.
%
%
%
%
Then, the Yukawa matrix $y_{ij}$ is rank 1, and thus,
only the third generation fermions can acquire masses by Higgs
mechanism.
%
When the family is replicated in the same torus,
the Yukawa matrix turns out to be diagonal and no mixing arises when
there is only one Higgs field.
Several discussions exist describing this property
\cite{Chamoun:2003pf,Abel:2003yh,Kitazawa:2004ed}.

The rank 1 property is an interesting feature
of the intersecting D-brane models as we will see in the subsequent sections.
In this section, we will suggest a possibility of raising  the rank.


\begin{figure}[t]
 \center
 \includegraphics[width=7cm]{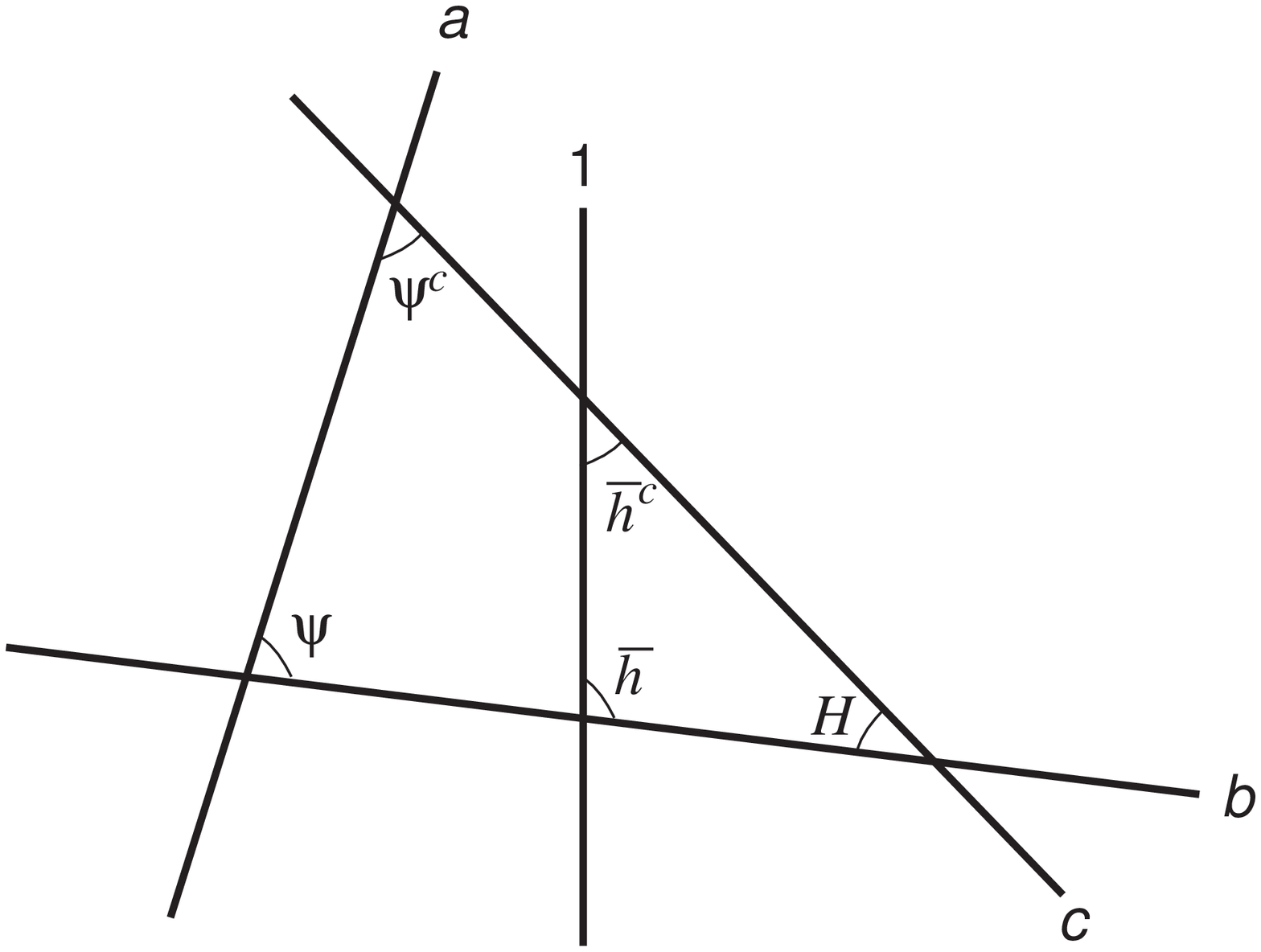}
 \caption{A sketch of intersections.  }
\label{schetch01}
\end{figure}

Let us suppose that the $USp$ brane passes across the triangle formed
by the $abc$ branes as shown in Figure 1.
Then the Higgs coupling $H \bar h \bar h^c$ arises in the superpotential
along with the Yukawa coupling $y\, \psi \psi^c H$.
As was calculated in \cite{Cvetic:2003ch}, from the quadrangle $ab1c$,
a four-fermi interaction
$y^\prime \psi \psi^c \tilde{\bar h} \tilde{\bar {h^c}}/M_s^2$
can arise which corresponds to a higher order term in the K\"ahler potential.
In non-supersymmetric models,
the four-fermi interaction can produce a Yukawa coupling by a loop diagram
\cite{Abel:2003yh}.
In supersymmetric models, 
such a loop effect is cancelled by a bosonic loop through the interaction
$y^\prime \psi \psi^c F_{\bar h}^* \bar h^c{}^\dagger$,
where $F_{\bar h}^* \sim H \bar h^c$.
However, in our model construction, $\bar h^c$ can acquire a
vacuum expectation value, and thus a new Yukawa coupling,
$\displaystyle y^\prime \frac{\langle \bar h^c \rangle^2}{M_s^2} \psi \psi^c H$,
can be produced from a K\"ahler potential term directly.
Contrary to a three-point function, the coupling $y^\prime$ does not factorize in general
 since it involves a four-point function~\cite{Cvetic:2003ch,Abel:2003yh}.
This is due to the fact that there are more parameters in the four point
function and only in certain limiting cases one obtains the factorizable result.
Therefore, due to the existence of the
$USp$ branes, which may be needed to cancel the RR tadpole,
the fermions of 1st and 2nd generations can acquire masses
both in supersymmetric and non-supersymmetric scenarios.
When the massless chiral Higgs fields are $h$ and $h^c$ instead of $\bar h$ and $\bar h^c$, the superpotential term
$\psi \psi^c h h^c$ will be produced and will contribute to fermion masses.
We note that the structure is consistent when we consider the confining phase of $USp$ gauge group. In that case, the
second generation masses are suppressed by the ratio of the confining scale and the string scale. When the beta function
of $USp$ group is negative, the mass hierarchy can be naturally explained.

%

We note that the Higgs bidoublet may acquire a large mass
through the $H \bar h \bar h^c$ coupling when $\bar h^c$ gets
a vacuum expectation value around the string scale.
However, since there exists $SU(2)_L$ singlets in the $bb^\prime$
sector, one can obtain bilinear masses of $H$ and $\bar h$
and thus a linear combination can be made to be light.
In any case, such mixings may be needed to break the
up-down quark mass symmetry arising from the existence of an $SU(2)_R$
symmetry.



\section{Properties of ``almost rank 1" Yukawa matrix}

In the previous section, we discussed the construction of the Pati-Salam model.
In the model, the Yukawa coupling can be a rank 1 matrix plus small
contributions from higher order terms.
We will call such a Yukawa matrix as ``almost rank 1 matrix".
One may think that there is no prediction once the higher order terms are added.
However, such an almost rank 1 matrix gives us several
 qualitative features for the fermion masses and mixings.
For example, the masses of 1st and 2nd generations can be hierarchically
smaller than the masses of the 3rd generation.
For the mixings, there are interesting qualitative properties as well.
In this section, we will see the properties of
fermion mixings in a general framework which does not depend very much
on the details of the model.

%

The Yukawa matrices for quarks and leptons are approximately given as
rank 1 matrices $y_{ij} = x_i^L x_j^R$.
For simplicity, let us consider the symmetric matrix $x^L = x^R$.
It can be easily extended to the case of non-symmetric matrices.
The rank 1 matrix is written as
\begin{equation}
Y_0 = \left(\begin{array}{c}
c \\ b \\ a
\end{array}
 \right) \left(\begin{array}{ccc} c & b & a \end{array}\right)
= \left(\begin{array}{ccc}
c^2 & b c & a c \\
b c & b^2 & a b \\
a c & a b & a^2
\end{array}
\right).
\end{equation}
The parameters $a,b,c$ can be made real, and are generically $O(1)$ parameters.
We obtain a useful unitary matrix to diagonalize the rank 1 matrix\,:
\begin{equation}
U_0 Y_0 U_0^T = {\rm diag} (0, \ 0, \ a^2+b^2+c^2),
\end{equation}
\begin{equation}
U_0 = \left(
\begin{array}{ccc}
\frac{b}{\sqrt{b^2+c^2}} & - \frac{c}{\sqrt{b^2+c^2}} & 0 \\
\frac{ac}{\sqrt{b^2+c^2}\sqrt{a^2+b^2+c^2}} &
\frac{ab}{\sqrt{b^2+c^2}\sqrt{a^2+b^2+c^2}}
& - \frac{\sqrt{b^2+c^2}}{\sqrt{a^2+b^2+c^2}} \\
\frac{c}{\sqrt{a^2+b^2+c^2}} & \frac{b}{\sqrt{a^2+b^2+c^2}} &
\frac{a}{\sqrt{a^2+b^2+c^2}}
\end{array}
\right).
\end{equation}
It is useful to parameterize as
\begin{equation}
a = \sqrt{y} \cos \theta_a, \quad
b = \sqrt{y}\sin \theta_a \cos \theta_s, \quad
c = \sqrt{y} \sin \theta_a \sin \theta_s \,.
\end{equation}
%
%
It is important to note here that there are only
two independent angles in the unitary matrix.
Since two of the eigenvalues are zero,
the corresponding eigenvectors can be rotated to a linear combinations
of the 1st and 2nd row vectors in the unitary matrix. 
However, when the rank 1 matrix is perturbed to raise the rank,
it can be easily checked that this basis is useful to
perturb in the limit where the 1st generation is massless.



Let us consider ``almost rank 1" Yukawa matrices for quarks as
\begin{equation}
Y_u = Y_0 + \tilde Y_u, \qquad Y_d = Y_0 + \tilde Y_d \,.
\label{quark-Yukawa}
\end{equation}
The $\tilde Y_{u,d}$ are  perturbation matrices.
Let us start on a basis where $\tilde Y_d$ is diagonal,
\begin{equation}
\tilde Y_d = {\rm diag} (\epsilon_1, \epsilon_2, \epsilon_3).
\end{equation}
To obtain the hierarchical quark masses,
 we need to assume $\epsilon_1, \epsilon_2 \ll \epsilon_3 \ll a,b,c$.
The eigenvalues of the down-type Yukawa matrix are approximately
\begin{equation}
y_b \simeq y,
\quad
y_s \simeq \sin^2 \theta_a^q \epsilon_3,
\quad
y_d \simeq \cos^2 \theta_s^q \epsilon_1 + \sin^2 \theta_s^q \epsilon_2 +
\cot^2 \theta_a^q \epsilon_1 \epsilon_2/\epsilon_3.
\end{equation}
To make the basis clear, we have attached
the superscript $q$ to  $\theta_a$ and $\theta_s$.
In this basis, $\tilde Y_u$ is not necessarily diagonal,
but it is reasonable to assume that $\tilde Y_u$ is almost aligned to $\tilde Y_d$.


Defining the unitary matrix $V_u, V_d$ such that
$V_u Y_u V_u^T$ and $V_d Y_d V_d^T$ are diagonal,
we obtain the CKM matrix as $V_{\rm CKM} = V_u V_d^\dagger$.
Although $V_u$ and $V_d$ include large mixing angles in $U_0$,
such large mixings are canceled out in the CKM matrix
because the left-handed rotation $U_0$ is common in $Y_u$ and $Y_d$.
 We note that even if we do not have the left-right symmetry,
 the left-handed rotation is common in the formulation of the model,
$Y_0^{(u,d)}{}_{ij} = x^L_i x_j^{R (u,d)}$.
One can define unitary matrices $\tilde V_{u,d}$ such that
$V_{u,d} = \tilde V_{u,d} U_0$ and
$V_{\rm CKM} = \tilde V_u \tilde V_d^\dagger$.
The $\tilde V_d$ is the diagonalizing matrix of $U_0 Y_d U_0^T$:
\begin{equation}
U_0 Y_d U_0^T \simeq
\left(
\begin{array}{ccc}
\cos^2 \theta_s^q \epsilon_1 + \sin^2\theta_s^q \epsilon_2 &
\frac12 (\epsilon_1-\epsilon_2) \cos \theta_a^q \sin2\theta_s^q &
\frac12 (\epsilon_1-\epsilon_2) \sin \theta_a^q \sin2\theta_s^q \\
\frac12 (\epsilon_1-\epsilon_2) \cos \theta_a^q \sin2\theta_s^q &
 \sin^2 \theta_a^q \epsilon_3 & -\frac12 \sin2\theta_a^q \epsilon_3 \\
\frac12 (\epsilon_1-\epsilon_2) \sin \theta_a^q \sin2\theta_s^q  &
 - \frac12\sin2\theta_a^q \epsilon_3 & y+ \cos^2\theta_a^q \epsilon_3
\end{array}
\right).
\end{equation}
Since the up-type quarks are more hierarchical than the down-type ones,
one can expect that
$V_{\rm CKM} \simeq \tilde V_d^\dagger$.
We then obtain $V_{cb}\simeq \frac12 \sin2\theta_a^q \epsilon_3/y$,
or,
\begin{equation}
V_{cb} \simeq \cot\theta_a^q \ \frac{m_s}{m_b}\,.
\end{equation}
Since $a,b,c$ are expected to be $O(0.1)$ parameters, we obtain $V_{cb} \sim m_s/m_b$ as a string scale relation which
is in agreement with experiments \cite{Eidelman:2004wy}. We have little more flexibility to fit the other two angles.
Now we set the famous empirical relation $V_{us} \simeq \sqrt{m_d/m_s}$ as input. In order to do so we assume $(U_0 Y_d
U_0^T)_{11}\simeq 0$, which leads to $\epsilon_1 \simeq -\tan^2\theta_s^q \epsilon_2$. Using this relation we obtain
$V_{us}\simeq \cos\theta_a^q/\sin^{2}\theta_a^q 
 \tan\theta_s^q \epsilon_2/\epsilon_3$,
$V_{ub} \simeq \sin\theta_a^q \tan\theta_s^q \epsilon_2/y$.
We finally  obtain the following relation
\begin{equation}
V_{ub} V_{cb} \simeq \left(\frac{m_s}{m_b}\right)^2 V_{us}\,,
\end{equation}
which is again in good agreement with experiments.
The Kobayashi-Maskawa phase can be derived from a phase of $\epsilon_3$.



Next, let us go on to the leptonic sector.
If type I seesaw contribution (i.e. $M_\nu^D M_R^{-1} (M_\nu^D)^T$) is dominant,
the large mixings in $Y_0$ are
canceled between the charged-lepton and the neutrino Dirac Yukawa couplings,
in the same way as it happens in the CKM matrix.
However, if the type I contributions are suppressed due to a large
right-handed Majorana mass scale, the large mixings can directly appear in general.
We therefore consider the case where $M_L$ dominates
in the light neutrino mass formula, Eq.(\ref{seesaw}).
We start in the basis where the light neutrino mass matrix is diagonal.
The charged-lepton Yukawa matrix is given as
\begin{equation}
Y_e = Y_0^l + \tilde Y_e.
\end{equation}
We note that $Y_0^l$, given in the above basis,
may be different from the $Y_0$ where $\tilde Y_d$ is diagonal even if
the $SU(4)_c$ unification is exact.
However, $a^l,b^l,c^l$ in $Y_0^l$ are all $O(1)$ parameters in general in this
basis.
 (We have attached the superscript $l$ to avoid any confusion).
The $\tilde Y_e$ is not necessarily diagonal in this basis,
but it may be reasonable that the $\tilde Y_e$ is close to diagonal
since it is hierarchical.

%

The Maki-Nakagawa-Sakata-Pontecorvo (MNSP) matrix
for neutrino oscillation is given in the basis as
$U_{\rm MNSP} = V_e^*$ where $V_e Y_e V_e^T$ is diagonal.
The unitary matrix $\tilde V_e$ is defined as
 $V_e = \tilde V_e U_0^l$ and it is the
diagonalization matrix of $U_0^l Y_e U_0^l{}^T$.
The matrix $\tilde V_e$ can be parameterized as
\begin{equation}
\tilde V_e = \left(
\begin{array}{ccc}
c^e_{12} c^e_{13} & s^e_{12} c^e_{13} & s^e_{13}e^{i \delta} \\
-s^e_{12} c^e_{23} - c^e_{12}s^e_{23} s^e_{13}e^{-i \delta} &
c^e_{12} c^e_{23} - s^e_{12} s^e_{13} s^e_{23} e^{-i \delta} &
c^e_{13} s^e_{23} \\
s^e_{12} s^e_{23}- c^e_{12} c^e_{23} s^e_{13} e^{-i \delta} &
-c^e_{12} s^e_{23} -s^e_{12} s^e_{13} c^e_{23} e^{-i \delta} &
c^e_{13} c^e_{23}
\end{array}
\right)
\left(\begin{array}{ccc}
1 & & \\
& e^{-i \alpha} & \\
& & e^{-i\beta}
\end{array}
\right).
\end{equation}
Then 13 element of MNSP matrix, $U_{e3}$, can be calculated as
\begin{equation}
U_{e3} = \cos \theta_a^l e^{i (\beta-\delta)} s_{13}^e
- \sin\theta_a^l e^{i\alpha} s_{12}^e c_{13}^e\,.
\end{equation}

The charged-lepton masses are hierarchical,
and thus we expect that all three mixing angles in $\tilde V_e$ are small in
the same way as $\tilde V_d$.
Since the up-type Yukawa is more hierarchical rather than the down-type one,
we expect the relation $V_{\rm CKM} \simeq \tilde V_d^{\dagger}$ approximately.
When we consider a quark-lepton unification, we can
expect that $\tilde V_e \sim \tilde V_d \sim V_{CKM}^\dagger$.
So, we neglect $s_{13}^e$, $s_{23}^e$, which are much
smaller than $s_{12}^e$. Then we obtain three mixing angles,
$\sin \theta_{13} \equiv |U_{e3}|$,
$\tan \theta_{\rm sol} \equiv |U_{e2}/U_{e1}|$,
and $\tan \theta_{\rm atm} \equiv |U_{\mu 3}/U_{\tau 3}|$,
 for the neutrino oscillation approximately,
\begin{equation}
 \sin\theta_{13}
 \simeq \sin\theta_a^l s_{12}^e\,,
\end{equation}
\begin{equation}
\tan \theta_{\rm atm} \simeq \tan \theta_a^l c_{12}^e\,,
\end{equation}
%
%
\begin{equation}
\sin^2\theta_{\rm sol} \simeq \sin^2 \theta_s^l
\left(1- 2\cot \theta_s^l \cos\theta_a^l t_{12}^e \cos \alpha +
\frac{\cos 2\theta_s^l}{\sin^2 \theta_s^l} \cos^2 \theta_a^l t_{12}^{e2}\right),
\label{solar}
\end{equation}
where $t_{12}^e = s_{12}^e/c_{12}^e$.
We note that an interesting approximate relation,
\begin{equation}
\theta_{\rm sol} \sim \theta_s^l \pm \theta_{13} \cot \theta_{\rm atm} \cos \alpha\,,
\end{equation}
is satisfied. 
The Jarskog invariant of the MNSP matrix can be calculated as
\begin{equation}
J_{\rm MNSP} \simeq \frac18 \sin 2 \theta_{12}^e \sin 2 \theta_a^l \sin 2\theta_s^l
\sin \theta_a^l \sin\alpha \,.
\end{equation}
Neglecting a small $s_{12}^e (= \sin\theta_{12}^e)$,  
we find that the phase $\alpha$ corresponds to the MNSP phase
approximately up to a quadrant.
Therefore, when CP violation in neutrino oscillation is maximal
(which corresponds to $\sin \alpha = \pm1$),
the solar mixing angle is almost same as $\theta_s^l$.

Let us assume, for example, that $\theta_a$ and $\theta_s$ are maximal (45 degree),
$\alpha=0$, $t_{12}^e>0$, and $s_{12}^e$ is the same as the Cabibbo angle
($s_{12}^e = 0.22$).
Then we find $\sin^2 2\theta_{\rm atm} \simeq 1$, $\tan^2\theta_{\rm sol}
\simeq 0.52$, $U_{e3} \simeq 0.15$ which are consistent with current experimental
data.

The 12 mixing $s_{12}^e$ may be smaller than the Cabibbo angle by a
 factor 1/3 because the muon mass is a factor of 3 larger than
the strange quark mass. So, $U_{e3}$ is expected to be in the range $0.05 - 0.15$.

Note that the bi-large mixing angles and a small $U_{e3}$
can be naturally obtained from the almost rank 1 charged-lepton Yukawa matrix.
The two angles $\theta_a^l$ and $\theta_s^l$ are due to the rank 1 matrix and thus
those
are generically large mixings.
On the other hand, $\theta_{13}$ is generated from perturbation matrix,
and thus, it is naturally small.
This qualitative feature does not depend on the details of the model.

\section{Geometrical interpretation of neutrino mixings}

In the previous section,
we have seen that patterns of the observed quark and lepton mixings
can be easily reproduced using the almost rank 1 Yukawa matrices.
Although there are no rigid quantitative predictions,
the qualitative feature is interesting especially for the neutrino mixings.
The solar and atmospheric mixings are generically large and $\theta_{13}$ is small.
The reason for $\theta_{13}$ mixing being small only can be explained geometrically
i.e. the left- and the right-handed families are replicated on different tori.
Further, we have seen that the solar and atmospheric mixings
are almost same as the two angles in a rank 1 Yukawa matrix.
So, these two mixings can be expressed by the Jacobi theta
function as a function of moduli parameters,
assuming that the mixings originating from left-handed Majorana mass matrix
are small.

\begin{figure}[t]
 \center
 \includegraphics[width=10cm]{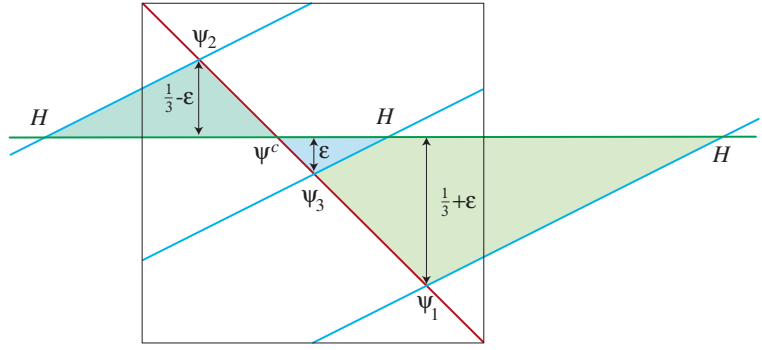}
 \caption{A sketch of brane intersections on a torus. }
\label{schetch02}
\end{figure}

As we have supposed to obtain an almost rank 1 Yukawa matrix,
$U(4)_c$ and $U(2)_R$  branes are intersecting once on the
torus where the left-handed matter is replicated.
So, let us assume that the intersecting numbers on a torus to be
$|I_{ab}^{(r)}|=3$, $|I_{ac}^{(r)}|=1$ and $|I_{bc}^{(r)}|=1$ as shown in Figure 2.
%
The ratio of the left-handed part $x^L_i=(c,b,a)$ of the rank 1 matrix,
 $y_{ij} = x^L_i x^R_j$,
is written as
\begin{equation}
a : b : c = \vartheta \left[
\begin{array}{c}
\varepsilon \\ 0
\end{array} \right](t) :
\vartheta \left[
\begin{array}{c}
-\frac13 +\varepsilon \\ 0
\end{array} \right](t) :
\vartheta \left[
\begin{array}{c}
\frac13 + \varepsilon \\ 0
\end{array} \right](t) \,.
\end{equation}
The moduli parameter $\varepsilon$ represents a shift of the D-brane as shown in
Figure 2, and $t= 3 A/\alpha^\prime$
where $A$ represents the K\"ahler structure of the torus. We neglect Wilson line
phase for simplicity. When $t$ is
large, which corresponds to a weak coupling limit, the ratio is determined by the
area of each triangle forming by
$\psi_i \psi^c H$. On the other hand, when $t$ is small, which corresponds to the
strong coupling limit, the
contributions from the triangles of larger sizes are not negligible. The shift
parameter $\varepsilon$ corresponds to
the vertical distance to the branes from the intersections where the left-handed
matter fields replicate. The
$\vartheta$ function is periodic for $\varepsilon$ which can be in the range $[0,1)$
 in general. But we assume that
$-\frac16 \leq \varepsilon \leq \frac16$ to identify the closest intersection to
be the third generation. Further,
turning on the 1st and 2nd generations for $\varepsilon<0$, we get
$0 \leq \varepsilon \leq \frac16$.
In this range of
$\varepsilon$, we obtain $a\geq b\geq c$. The angles $\theta_s$ and $\theta_a$ are
calculated as
 functions of
$\varepsilon$ and $t$, and they are plotted in the Figure 3 for different values
of $\varepsilon$. As one can see in the
figure that both mixings can be maximal.

Let us see the geometrical meaning of the behaviors of two mixings.
 At first, consider the case $\varepsilon = 0$ which
means that three branes are intersecting at one point in a torus.
In this case, $b=c$ for any $t$, and then $\tan\theta_s = c/b = 1$.
When $\varepsilon \neq 0$, the triangle $\psi_1 \psi^c H$
becomes larger than $\psi_2 \psi^c H$ and thus $b>c$.
Therefore, when $\varepsilon$ increases, $\theta_s$ is getting smaller.
Another typical case for the shift parameter is $\varepsilon = \frac16$.
In this case, the triangle areas for $\psi_3$ and $\psi_2$ are same and thus $a=b$.
Since $\tan\theta_a = \sqrt{b^2+c^2}/a$, the $\theta_a$ angle
becomes maximal when $c$ (and therefore $\tan\theta_s$)
gets exponentially damped for large $t$.

\begin{figure}[t]
 \center
 \includegraphics[width=8cm]{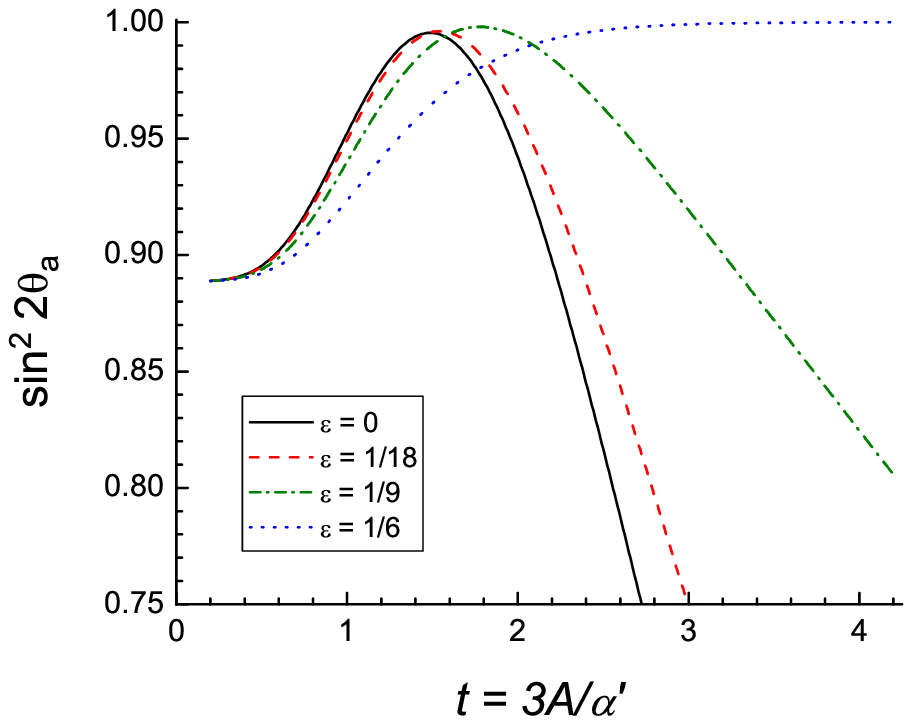}
%
 \includegraphics[width=8cm]{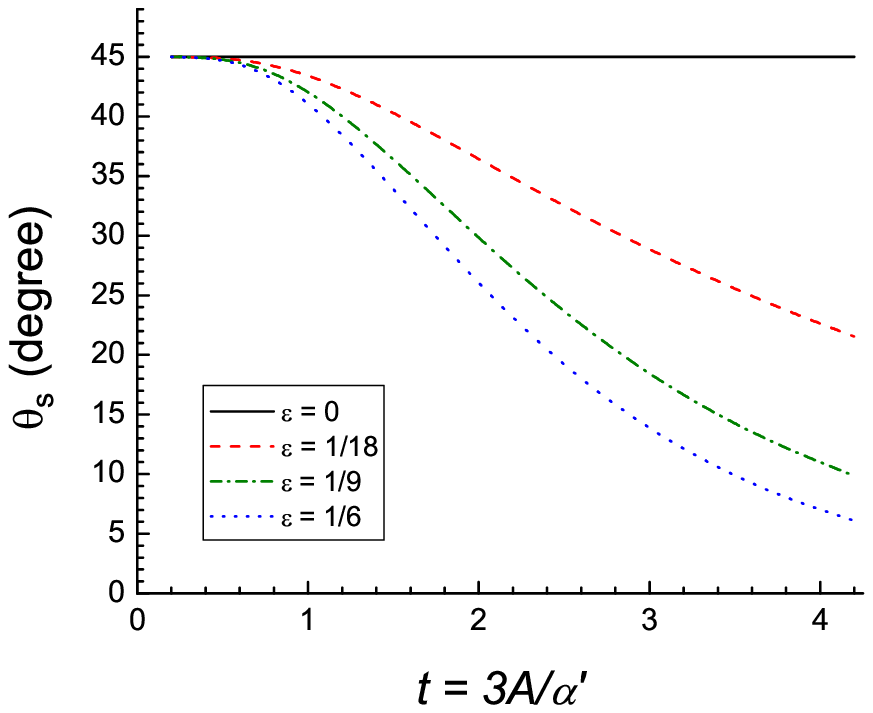}
 \caption{Plots for $\sin^2 2\theta_a$ and $\theta_s$ as functions of $\varepsilon$ and $t$.
The experimental data for atmospheric and solar mixings are
$\sin^2 2\theta_{\rm atm} > 0.92$ at 90\% CL and $\theta_{\rm sol} = (32 \pm 3)^{\rm o}$.}
\label{plot}
\end{figure}

Since in the weak coupling limit (large $t$),
the couplings are given as $e^{-k {\rm Area}}$.
One can easily see 
that the ratios $c/b$ and $b/a$ are exponentially
damped in the weak coupling direction 
except for the two cases $\varepsilon = 0$ (for $c/b$), $\frac16$ (for
$b/a$). In the strong coupling limit, all the triangles shrink and
thus $a=b=c$. Thus, in the strong coupling limit, we have
$\tan\theta_a = \sqrt2$ and $\tan \theta_s = 1$ independent of
$\varepsilon$. In this limit, however, due to the instanton
corrections, the Yukawa couplings are blowing up and the effective
theory is not reliable. When $t$ increases to $t \agt 1$, the Yukawa
couplings become $O(1)$ and the field theory can be in a perturbative
region. In this region, $\theta_a$ always has the maximal mixing
solution and $\theta_s$ starts to descent from  the maximal value
except for $\varepsilon =0$.

The mixing angles $\theta_{a,s}$ can be modified to $\theta_{a,s}^l$
due to the mixing angles in the light neutrino Majorana mass matrix.
But, one can expect that the contribution from the Majorana matrix is small
$\theta_{a,s}^l \simeq \theta_{a,s}$.
We see that the solar and the atmospheric mixings can then be given by the moduli
parameters.
 The observed qualitative features for neutrino oscillations,
where the atmospheric mixing is almost maximal and the solar
mixing is not maximal but large,
can be interpreted geometrically when $t \sim 2$. 
This feature holds even for different intersection numbers.

 We have shown in the previous section that
the atmospheric mixing is  almost the same as $\theta_a^l$ while the solar mixing
can be modified as in Eq.(\ref{solar})
by $\theta_{13}$ and the MNSP phase. The calculated value of $\theta_s$
($\varepsilon \neq 0$) is consistent with the
observed solar mixing even if CP violation is maximal or small $\theta_{13}$ mixing. As shown in
 section 4, the observed data (for $\theta_{\rm sol}$ and $\theta_{\rm atm}$) are
 consistent with bi-maximal mixing $\theta_a = \theta_s = 45^{\rm o}$
 (which corresponds to
$\varepsilon =0$ and $t = 1.5$) when $U_{e3}$ is close to current
experimental bound ($\theta_{13} < 10^{\rm o}$ at 99\%
CL) and no CP violating phase in neutrino oscillation. We can distinguish
two situations $\varepsilon =0$ or
$\varepsilon \neq 0$ in the future long baseline experiments to measure
$U_{e3}$ and the MNSP phase \cite{Anderson:2004pk}.

\section{Conclusion}

In this paper, we have studied Yukawa coupling structures
in the intersecting D-brane models with the Pati-Salam gauge
symmetry with extra $U(1)$ symmetries.
The Yukawa matrices are almost rank 1 when the left- and right-handed matters are
replicated on different tori.
Because of the existence of $USp$ branes,
four-point interaction can appear and the rank
of the Yukawa matrices goes up due to the perturbation effects
to the rank 1 matrices.
With the almost rank 1 matrices,
the observed quark and lepton mixings can be naturally reproduced.
Especially, for the neutrino mixings,
the bi-large and a small $\theta_{13}$ mixings 
can be naturally realized.
Further, in the quark sector, $V_{cb}$ is naturally close to the strange/bottom
quark mass ratio, and there exists a
simple relation among the CKM mixing angles and a quark mass ratio. In the
neutrino sector, the important prediction is
that  $U_{e3}$ is related to the 12 mixing in charged-lepton sector. Consequently,
if the
quark-lepton unification is
 realized simply, we predict $U_{e3} \simeq 0.05-0.15$ and
almost the entire range of this prediction can be tested at future
long baseline experiments \cite{Anderson:2004pk}.
 We have also studied the geometrical meaning of the fact that the atmospheric
mixing is almost maximal and the solar mixing is large but not
maximal. This feature can arise when the Yukawa coupling is in the
perturbative region. The results of the future long baseline
experiments will be useful to shed light on
 geometrical interpretations.

We emphasize that the properties of fermion mixings which is
reproduced from ``almost rank 1" Yukawa matrices are model independent.
These properties do not depend on how the rank of Yukawa matrices is raised.
The crucial assumption is that the Yukawa coupling matrices are
rank 1 plus small perturbations and the seesaw neutrino masses are type II dominant.

The ``almost rank 1" matrices may be constructed in usual particle field theories,
for example, by using Froggatt-Nielsen mechanism \cite{Froggatt:1978nt}
with an appropriate flavor symmetry.
Discrete flavor symmetry can also construct the rank 1 matrices.
However, the rank 1 Yukawa matrices in the intersecting D-brane are not
originating from symmetrical reason but from the geometrical configuration of
the matter representation on tori.
It is interesting that such patterns of fermion mixings in nature
are naturally derived from a simple assumption in the context of string theory.
The results can encourage us to understand the variety of
quark and lepton masses and mixings in fundamental theories.

\section*{Acknowledgments}
We thank M. Cvetic for valuable discussion.

\end{document}